\documentstyle[12pt]{article}
\makeatletter
 
 \@addtoreset{equation}{section}
\makeatother
\begin{document}
\newcommand{\be}{\begin{equation}}
\newcommand{\ee}{\end{equation}}
\newcommand{\vth}{\vspace{3mm}}
\newcommand{\bec}{\begin{center}}
\newcommand{\ec}{\end{center}}
\newcommand{\befl}{\begin{flushleft}}
\newcommand{\efl}{\end{flushleft}}
\newcommand{\hed}{\enskip \quad}
\newcommand{\dfrac}[2]{%
        \frac{\displaystyle{#1}}{\displaystyle{#2}}}
\newcommand{\mapdown}[1]{\Big\downarrow
   \rlap{$\vcenter{\hbox{$\scriptstyle#1\,$}}$ }}
\newcommand{\mapright}[1]{%
  \smash{\mathop{%
   \hbox to 0.5cm{\rightarrowfill}}\limits^{#1}}}
\newcommand{\mapleft}[1]{%
      \smash{\mathop{%
        \hbox to 0.5cm{\leftarrowfill}}\limits^{#1}}}
\begin{flushright}
{TMUP-HEL-9902}
\end{flushright}
\vspace*{1.7cm}
\bec
\LARGE{ On Matrix Description of IIA Center of Mass}\\[1.5cm]
\normalsize{ Noriaki Ano}$~ ^\dagger$
\footnote[0]{\noindent $\dagger$~E-mail: ano@phys.metro-u.ac.jp}\\
\normalsize{\it Department of Physics, Tokyo Metropolitan 
University,}\\
\normalsize{\it Minamioosawa, Hachiooji, Japan}\\
\ec
\normalsize
\vspace{1.0cm}
\bec Abstract \ec
The correspondence between the ground states
of the BFSS matrix model and the type IIA string
is investigated through the 11th direction.
We derive the type IIA string from 11$D$ supermembrane
wrapped around the 11th direction of periodicity
without a procedure of the double dimensional reduction,
but by taking the limit $R_{11}\rightarrow \infty$.
It is shown that the center of mass of
this string as a function on the 11th coordinate $X^{11}$
has the same matrix
description as the BFSS matrix model
in the flat direction.
The fact shows that
the matrix model at the strong string coupling
in the flat direction of large distance regime
is directly connected with
the 11$D$ supermembrane theory in the transverse light-cone frame.

\vspace{2.5cm}
\begin{flushleft}
\small{Feb. 1999}
\end{flushleft}

\newpage
\section{\rm Introduction}
The BFSS matrix model \cite{bfss} is
regarded as a suitable 
candidate for
realizing the eleven dimensional {\it M} theory.
The most solid ground for this conjecture
is the fact that the model contains the excitations of
both the basic 11-dimensional supergravity multiplet
with 256 degrees of freedom, and
large classical supermembrane in the light-cone frame.
The contents of 256 degrees of freedom are
44 gravitons, 84 components of a 3-form field and
128 gravitinos, i.e., the supergraviton.
The supergraviton degrees of freedom
are equivalent to the lowest states of type IIA string
which can be obtain
by the dimensionally reduced
supermembrane wrapped around the eleventh direction
from the 11$D$ view point\cite{duff}\cite{townsend}.
Moreover,
we know that by compactifing
the ninth transverse direction
(i.e., the 11-th direction of {\it M} theory
with an 11-9 flip) on a circle
of small radius $R_{9}$,
the matrix model of weak string coupling
can be reduced to a multiple of the light-front
type IIA string on the dual torus,
on the assumption that all the
matrices are defined all over the
ninth direction \cite{dvv-bs},
based on the works by Motl \cite{motl} and
by Sethi and Susskind \cite{sethi-susskind}.
The matrix model in this case is modified as
noncommuting membrane theory
extended in the ninth direction.
We can see that the
perturbatively
interacting type II string spectrum
which the matrix model may originally contains, 
become visible through this compactification procedure.

Clearly, these features indicate that the matrix model is
intrinsically related to the IIA string.
Originally the matrix model conjecture
is motivated by this relation.
In fact, the model is founded upon
the $N$ D0-branes mechanics with
nonperturbative R-R charge of the type IIA string.
On the other hand, the correspondence between the matrix model
and the eleven dimensional $M$ theory is not very strict.
That is
because the matrix model formulation is based only on the weakly
coupled IIA string theory with small radius $R_{11}$ of 
compactified 11th direction.
In case of the limit $R_{11}\rightarrow \infty$,
i.e., the strong string coupling regime,
it seems not sure that this conjecture could be correct for
describing the nature of the 11 dimensional $M$ theory.

In this paper,
we will show that
the matrix model in the flat direction
has a relation with the 11 dimensional
supermembrane wrapped around
the 11th direction in $R_{11}\rightarrow \infty$ limit,
or in string theory words, the strong coupling regime.
In the ordinary Kaluza-Klein (KK) mechanism, 
the radius $R$ of some periodic coordinate
is set to small to obtain physics in the transeverse space-time.
Thus, it is found that the heavy KK nonzero-modes
turn out to decouple from
the massless sector.
We instead discuss the large $R$ limit.
Taking this limit leads us to the massless KK
nonzero-modes.
All the KK modes then
correspond to the massless degrees of freedom.
We can see that the massless nonzero-modes in the
large $R$ limit could not be neglected.
Henceforce, we will find that the 11$D$ supermembrane in this limit
has a matrix description.

The paper is organized as follows.
In the next section {\bf 2},
the 11$D$ supermembrane wrapped around the 11th direction
is considered.
It is shown that
by taking the limit $R_{11}\rightarrow\infty$,
the type IIA string is obtained without using
the double dimensional reduction.
It is found that
this IIA string is on
the periodic coordinate $X^{11}$.
In section {\bf 3},
we study the center of mass of the $X^{11}$-dependent type IIA stirng
obtained in section {\bf 2}, which corresponds to the lowest
states of the type IIA string.
In order to deal with the physical degrees of freedom,
we take the light-cone gauge.
Furthermore, we next make the reduction
of one of the unwrapped coordinates
on the membrane world-volume to obtain the action of the center of mass
of the $X^{11}$-dependent type IIA string.
We are then lead to the action describing the
center of mass on loop.
In section {\bf 4},
the matrix description of the $X^{11}$-dependent action
obtained in the previous section
is presented through the Fourier expansion
with respect to the periodic $X^{11}$ direction.
In the final section {\bf 5},
some discussions thought to be useful for justifing the results
are given.
\section{\rm $X^{11}$-Dependent Type IIA String}
In the KK-compactification of certain space direction with
its small radius $R$,
the KK nonzero-modes become very heavy
and decouple from the massless zero-mode in the result.
Thus, it is allowed to pick up only the massless
KK zero-mode in low energy theory.
This procedure means the dimensional reduction
in general.
We know that
the type IIA strings can be derived from 11$D$ supermembrane
wrapped around the 11th direction
by using the double dimensional reduction along
this direction \cite{duff}.
In this section,
if we are concerned with only
the elimination of the mass terms
of the KK nonzero-modes,
we will see that there could be another way to 
eliminate all the mass terms without
decoupling the KK nonzero-modes.
We must remind that the procedure of
the double dimensional reduction amounts to
pick up the type IIA string as the KK zero-mode
which decouples from the KK nonzero-modes
in the small $R_{11}$ limit.
\\

\noindent
\underline{{\it 2-1.}\enskip IIA String from Supermembrane}

\noindent
Let us first recall the derivation of the type IIA string from 
the supermembrane wrapped around the compactified eleventh 
direction
via double dimensional reduction
of the world volume theory in 11$D$
space-time
to the world sheet theory in 10$D$ space-time \cite{duff}.
Let us suppose that
the world volume coordinates are $\tau,\sigma^1$ and 
$\sigma^2$.
The coordinate $\sigma^2$ is set identical to
the compactified eleventh space direction
as a gauge choice; $\sigma^2 =X^{11}$.
The 11$D$ supermembrane action ${\cal S}^{SM}$
without a 3-form gauge field coupling
can be presented as follows;
\be
{\cal S}^{SM}=T_{SM}
  \int d\tau d\sigma^1 dX^{11}
  \sqrt{\mid \mbox{det}~g^{SM}_{\hat{a}\hat{b}}  \mid },
\label{supermembrane}
\ee
where $g^{SM}_{\hat{a}\hat{b}}$
is an induced metric
on the world-volume of the supermembrane(SM), and
the factor $T_{SM}$ denotes the membrane tension;
$T^{SM}=1/(2\pi)^2l_{p}^3
=1/(2\pi)^2l_{s}^3g_{s}$.
The letters $\hat{a},\hat{b}$ denote the world-volume
coordinates $(\tau, \sigma^1, X^{11})$.
The induced metric
$g^{SM}_{\hat{a}\hat{b}}$ can be described by
using super-space coordinates of supermembrane
$Z^{\hat{M}}=(X^m,X^{11},\theta^{\mu})~(m=0,1,
\cdot\cdot\cdot,9~\mbox{and}~\mu = 1,\cdot\cdot\cdot,32)$
and the 11$D$ supergravity background metric
$G_{\hat{M}\hat{N}}$;
\be
g^{SM}_{\hat{a}\hat{b}}=\partial_{\hat{a}}Z^{\hat{M}}
    \partial_{\hat{b}}Z^{\hat{N}}
                  G_{\hat{M}\hat{N}}.
\ee
The KK massive modes on the $\tau$-$\sigma^1$ world sheet
are eliminated by the dimensional reduction of $X^{11}$,
leaving only the KK massless zero mode on the world sheet.
This procedure can be yielded
by putting the following condition;
\be
\frac{\partial}{\partial X^{11}}~\mbox{( ALL FIELDS )}=0.
\label{condition-DR}
\ee
The condition (\ref{condition-DR})
which corresponds also to setting all fields
independent of $X^{11}$,
leads us to the reduction of $g^{SM}_{\hat{a}\hat{b}}$
to the metric $g^{GS}_{ab}$
on the $\tau$-$\sigma^1$ world sheet.
We can read the reduced $g^{GS}_{ab}$
by using the super-coordinates
$Z^{M}(\tau,\sigma^1)=(X^m,\theta^{\mu})$ as follows;
\be
g^{GS}_{ab}=\partial_{a}Z^{M}
    \partial_{b}Z^{N}
                  G_{MN},
\ee
where $G_{MN}$ is a reduced background metric from the
11$D$ supergravity.
We therefore reach at the GS-string (GS) action
of type IIA;
\be
{\cal S}^{GS}=2\pi R_{11} T_{SM}
  \int d\tau d\sigma^1
  \sqrt{\mid \mbox{det}~g^{GS}_{ab} \mid },
\ee
where $R_{11}$ is a radius of the compactified $X^{11}$ coordinate,
and the following condition on the string tension $T_{s}$
must be fulfilled;
\be
T_{s}=2 \pi R_{11} T_{SM}.
\label{m-s}
\ee
Consequently, the type IIA theory coupled with the 10$D$
supergravity background is obtained by
this procedure of the double dimensional reduction.\\

\noindent
\underline{{\it 2-2.}\enskip
       IIA String without Double Dimensinal Reduction}

\noindent
We have now arrived at a subtle, but rather essential point.
We would like to focus on eliminating the mass terms
of KK nonzero-modes.
We can see that there is another way
to vanish the mass of all the KK-modes
beside setting the condition (\ref{condition-DR}),
i.e., without vanishing the $X^{11}$ dependence.
As will be seen in the following context,
we will find that in this case, the KK nonzero-modes could not
decouple from the massless sector.

The periodic $X^{11}$ coordinate is an $S^1$
coordinate with its radius $R_{11}$,
and we naturally introduce a dimensionless angle parameter
$\tilde{\tau}$, which satisfies $X_{11}=R_{11}\tilde{\tau}$.
The differentiation associated with $X^{11}$ is written
by using $R_{11}$ as follows;
\be
\frac{\partial}{\partial X^{11}}=
\frac{1}{R_{11}}\frac{\partial}{\partial \tilde{\tau}}.
\ee
It is well-known that the mass
of the KK-modes is yielded from
this factor $1/R_{11}$, and that in the limit 
$R_{11}\rightarrow\infty$
the mass of all the KK-modes vanishes.
That is, all the KK-modes turn out to be the massless modes.
The very important point we would
like to emphasize is that in this limit,
all the terms operated by $\partial/\partial X^{11}$ vanish.
On the other hand, the terms which do not contain them
survive with no change.
Thus, we obtain the massless KK nonzero-modes
without putting the condition (\ref{condition-DR}).
In this procedure,
we see that the $X^{11}$ dependence
of all the fields is preserved in contrast to
the result by making the ordinary dimensional reduction,
that is, all the fields can be regarded as functions
defined on $X^{11}$.
Therefore, starting with the
supermembrane action (\ref{supermembrane}),
we then obtain the following action by taking the limit
$R_{11}\rightarrow \infty$;
\begin{eqnarray}
&&\lim_{R_{11}\rightarrow \infty}
{\cal S}^{SM}  = 
\int d\tilde{\tau}~{\cal S}^{GS}(\tilde{\tau}) 
\label{limiting-action}\\
&&{\cal S}^{GS}(\tilde{\tau}) =
R_{11} T_{SM}\int d\tau d\sigma^1
\sqrt{\mid \mbox{det}~g^{GS}_{ab}(\tilde{\tau}) \mid }.
\label{gs-action}
\end{eqnarray}
${\cal S}^{GS}(\tilde{\tau})$ describes 
the $X^{11}$-dependent type IIA GS-string.
In this case,
the mass of all the modes turns out to be very small.
We therefore see that the KK nonzero-modes could not
decouple from the massless sector.

\section{\rm IIA Center of Mass on Loop}
In order to consider the lowest states of 
the $X^{11}$-dependent type IIA string (\ref{gs-action}),
we next deal with the string center of mass.
Thus in the following procedure,
we need to take the light-cone gauge and the dimensional reduction
along the $\sigma^1$ coordinate on the world-sheet.
In case of the type IIA string,
the center of mass corresponds to the lowest state,
and this fact is in contrast to
the case of the type IIB string.
Before going ahead with our discussion,
let us now consider the lowest states of the GS-strings
to specify a type of the GS-strings, which has
supermultiplet of the string center of mass with 256
degrees of freedom.\\

\noindent
\underline{{\it 3-1.}\enskip Lowest States of Center of Mass}

\noindent
We first recall that the contents of the GS-string degrees of freedom 
corresponds to the {\it D}=10 super Yang-Mills multiplet, 
i.e., eight Bose states $\mid i ~ \rangle$ of
the vector representation ${\bf 8_v}$ of $spin(8)$ 
labeled by the letters {\it i, j}, 
and eight Fermi sates $\mid a ~ \rangle$ 
(or $\mid\dot{a} ~ \rangle$ in opposite chirality)
of the spinor representation ${\bf 8_s}$ 
(or ${\bf 8_c}$) of $spin(8)$ 
labeled by {\it a, b}
(or {\it \.a, \.b}).
The $spin(8)$ contents of type IIA massless multiplet is
given by the tensor product of two supermultiplets of
opposite chirality
which is described by 256 states;
128 Bose states $\mid i ~ \rangle\mid j ~ \rangle ~ \oplus
  \mid \dot{a} ~ \rangle\mid a ~ \rangle$ and 128 Fermi states
$\mid i ~ \rangle\mid a ~ \rangle ~ \oplus
  \mid \dot{a} ~ \rangle\mid j ~ \rangle$.
In case of type IIB massless multiplet,
the tensor product of two supermultiplets
is given by Fermi states of the same chirality;
128 Bose states $\mid i ~ \rangle\mid j ~ \rangle ~ \oplus
  \mid a ~ \rangle\mid b ~ \rangle$ and 128 Fermi states
$\mid i ~ \rangle\mid a ~ \rangle ~ \oplus
  \mid b ~ \rangle\mid j ~ \rangle$.
On the other hand,
the type IIA and IIB equations of motion
of supermultiplets of the center of mass
in the light cone frame are as follows;
\begin{eqnarray}
& \mbox{type IIA :} & \frac{\partial^2}{\partial \tau^2} ~ X^i 
(\tau)=0,\quad
\frac{\partial}{\partial \tau} ~ S^{1a} (\tau)=0, \quad
\frac{\partial}{\partial \tau} ~ S^{2a} (\tau)=0,
\nonumber \\
& \mbox{type IIB :}& \frac{\partial^2}{\partial \tau^2}~ X^i (\tau) 
=0, \quad
\frac{\partial}{\partial \tau} ~ S^{1(2)a} (\tau)=0,
\label{ab}
\end{eqnarray}
where $S^1$ and $S^2$ belong to different spinor representations
of ${\bf 8_s}$ and ${\bf 8_c}$.
These supermultiplets correspond to zero modes of the GS-strings
in each case.
Thus from Eq.(\ref{ab}),
we see that the IIA field contents of supermultiplet of
the center of mass is composed of
fields of three different representations of $spin(8)$, 
i.e., ${\bf 8_v}$, ${\bf 8_s}$ and ${\bf 8_c}$,
while in case of type IIB, supermultiplet of
the center of mass contains two different representations of 
$spin(8)$,
i.e., ${\bf 8_v}$ and ${\bf 8_s}$ (or, ${\bf 8_c}$).
Then we find that the supersymmetric lowest states composed of 
the above
supermultiplets of the center of mass are as follows;
in case of type IIA, 256 states which are 
identical to type IIA massless multiplet
(that is, 128 Bose states $\mid i ~ \rangle\mid j ~ \rangle ~ \oplus
  \mid \dot{a} ~ \rangle\mid a ~ \rangle$ and 128 Fermi states
$\mid i ~ \rangle\mid a ~ \rangle ~ \oplus
  \mid \dot{a} ~ \rangle\mid j ~ \rangle$),
and in case of type IIB, 16 states which are  
identical with type I massless multiplet 
(that is, 8 Bose states $\mid i ~ \rangle$ and
8 Fermi states $\mid a ~ \rangle$ (or $\mid \dot{a} ~ \rangle$)).
As a consequence, we find that only in case of type IIA,
the supermultiplet of the center of mass has 256 degrees
of freedom which is the same contents as the massless
multiplet of the 11$D$ supergravity.\\

\noindent
\underline{{\it 3-2.}\enskip Action of $X^{11}$-Dependent IIA Center of Mass}

\noindent
Let us here take the light-cone gauge and the dimensional reduction
on the $X^{11}$-dependent IIA string (\ref{gs-action}),
in order to study its lowest states which correspond to
the string center of mass.
We first take the light-cone gauge.
In this gauge, the world-sheet parameter $\tau$ is
usually described as 
\be
\tau~\sim~X^+/~l_s^2~p^+,
\label{rc}
\ee
where $X^+$ (and $X^-$) denotes the light-cone coordinate
\be
X^\pm =\frac{X^0 \pm X^{9}}{\sqrt{2}},
\ee
and $l_{s}$ is the string length.
Here, we are free to set both $X^+$ and $p^+$ independent
of $X^{11}$ together with taking the light-cone gauge,
although all fields can be thought dependent upon $X^{11}$
as discussed above.
It is natural to take this choice because $\tau$
is independent of $X^{11}$ in the above
equation (\ref{rc}).
The other light-cone coordinate $X^-$ can also be eliminated
due to the Virasoro constraint equations as usual.
Thus, we can obtain the reduced type IIA string
action in the light-cone gauge from the 11$D$ supermembrane action.

Let us next perform the dimensional reduction to obtain
the center of mass.
This procedure corresponds to reduce
the IIA string action (\ref{limiting-action})
along the $\sigma^1$ direction
to one dimensional action.
By taking this procedure, it is meant vanishing
of $\sigma^1$-dependence of all the fields.
The procedure of the reduction can be done by putting the
condition; $\partial /\partial \sigma^1 = 0$.
Because the world sheet parameter $\sigma^1$
is dimensionless, the integration
with respect to $\sigma^1$ turns out to yield $2\pi$ alone.

Finally, we obtain the following $\tilde{\tau}$-dependent
light-cone action of type IIA center of mass
in the flat background coming from the 11$D$ supergravity;
\begin{eqnarray}
\lefteqn{{\cal S}(\tilde{\tau}) = }\nonumber\\
& &\lambda \int dt d\tilde{\tau}
\left(
\left( \frac{\partial X^i (t,\tilde{\tau})}{\partial t} \right)^2
 + i\left( S^{1a}(t, \tilde{\tau})
\frac{\partial S^{1a}(t,\tilde{\tau})}{\partial t}+
S^{2a}(t,\tilde{\tau})
\frac{\partial S^{2a}(t,\tilde{\tau})}{\partial t}\right)
\right),\nonumber\\
& & \label{com}
\end{eqnarray}
where $S^{1}$ and $S^{2}$ are the 8 component spinors
labeled by the letter $a$,
and redefined as $S\rightarrow l_s \sqrt{p^+}
S$.
The letter $i$ runs from 1 to 8.
The coefficient $\lambda$ is read as follows;
\begin{eqnarray}
\lambda &=& 2\pi l_{s}^2 p^+ R_{11} T_{SM}\nonumber\\
&=& \frac{p^+ R_{11}}{l_{s}g_{s}}.
\label{lambda}
\end{eqnarray}
\section{\rm Matrix Description}
The BFSS matrix model is a description of
a system in the infinite momentum frame along
the longitudinal direction.
The situation corresponds to taking the small
$R_{11}$ limit.
This model contains {\it N} D0-branes
as well as very short open strings
attached to D0-branes.
Specifically under the large distance scale,
the string sector becomes very massive and decouple largely
from the parton sector.
In this distance scale, all the matrices are commutative.
We will show that the
BFSS matrix model in this large distance regime
can be derived from 11$D$ supermembrane
wrapped around the 11th
direction in the large $R_{11}$ limit.
The discussions on the consistency of resulting matrix
description with the BFSS model in the large $R_{11}$ limit
are given in the next section {\bf 5}. \\

\noindent
\underline{{\it 4-1.}\enskip IIA Center of Mass as Diagonal Matrices}

\noindent
Due to the periodicity of $\tilde{\tau}$,
we have the Fourier mode expansions of $X^i (t,\tilde{\tau})$,
$S^1 (t,\tilde{\tau})$ and $S^2 (t,\tilde{\tau})$.
Here we must set the reality conditions on $X^i$, $S^1$, $S^2$;
\be
X_{-n}^i = X_n^{i\ast},\quad S_{-n}^{1}=S_{n}^{1\ast},\quad
S_{-n}^{2}=S_{n}^{2\ast}, \quad (n>0).
\label{real1}
\ee
We are free to set the additional reality conditions on the modes
$X_n^i$, $S_n^1$, $S_n^2$ as follows;
\be
X_{n}^i = X_n^{i\ast},\quad
      S_{n}^{1}=S_{n}^{1\ast},\quad
S_{n}^{2}=S_{n}^{2\ast}, \quad (n>0).
\label{real2}
\ee
Therefore, we have the following mode expansions;
\begin{eqnarray}
X^i (t, \tilde{\tau}) &=& \frac{1}{\sqrt{2\pi}}
\sum_{n\geq 0}^{\infty} 
X_n^i (t)\cos n\tilde{\tau}
\nonumber \\
S^1 (t, \tilde{\tau}) &=& \frac{1}{\sqrt{2\pi}}
\sum_{n\geq 0}^{\infty} 
S_n^1 (t)\cos n\tilde{\tau}
\nonumber \\
S^2 (t, \tilde{\tau}) &=& \frac{1}{\sqrt{2\pi}}
\sum_{n\geq 0}^{\infty} 
S_n^2 (t)\cos n\tilde{\tau}
\label{mode}
\end{eqnarray}
In ordinary treatment of Fourier expansion,
the conditions which we need here are the reality conditions
(\ref{real1}) on the original functions $X^i(\tilde{\tau})$.
By setting the additional reality conditions (\ref{real2}),
it is meant that
these functions can be regarded as
the space coordinates in the scheme of the matrix model.

Let us present matrix description of the resulting
$\tilde{\tau}$-dependent action of the previous section.
We first substitute Eq.(\ref{mode}) into Eq.(\ref{com}),
and integrate the action with respect to $\tilde{\tau}$.
We then obtain the action;
\begin{eqnarray}
{\cal S}&= & \lim_{N\rightarrow \infty}
 \sum^N_{n=0} \left[ \lambda \int dt \left(
\dot{X}^i_n \dot{X}^i_n
 + i \left( S^{1a}_n \dot{S}^{1a}_n +  S^{2a}_n \dot{S}^{2a}_n\right)
\right)\right]\nonumber\\
& = & \lim_{N\rightarrow \infty}
  \lambda \int dt~\mbox{Tr}\{(\dot{X}^i\dot{X}^i) 
       + i \theta^T \dot{\theta}\}.
\label{matrixaction}
\end{eqnarray}
$X^i$ denotes $N\times N$ real matrix where the index $i$ runs
from 1 to 8,
and $\theta$ is a 16-component spinor which is a
direct product of two 8-component spinors $S^1$, $S^2$;
\begin{equation}
\theta = ( S^1, S^2 )^T.
\end{equation}

\noindent
\underline{{\it 4-2.}\enskip Correspondence with Matrix Model}

\noindent
Clearly, all these matrices $X^i$, $\theta$
do not have nonzero off-diagonal elements.
Henceforce, we can see that the direction described by
these matrices is flat, and that
the action (\ref{matrixaction}) is the same form as
the well-known BFSS model in the flat direction
of large distance regime, except for
matching of the number of components of $X^i$.
That is, the BFSS model has nine $X^i$'s, while
the action of (\ref{matrixaction}) contains eight $X^i$'s.
It seems that the difference is
superficial one, because both cases have the same degrees
of freedom, 256.
In the matrix model, 
this number comes from a representation
of the algebra of the 16 fermionic field $\theta$'s with $2^8$ 
components.

To study this difference between the two matrix descriptions,
let us remember the original theory upon which the BFSS 
model
is founded, that is to say, the $D=10$ super Yang-Mills 
theory.
All of what we need here is instead the conserved
energy-momentum
tenser $T^{\mu \nu}$ of $D=10$ Yang-Mills theory with no 
supersymmetry.
\be
T^{\mu\nu}\sim ~ -\mbox{Tr}~F^{\mu\alpha}F^\nu_{\enskip \alpha}
      + 
\frac{1}{4}\mbox{Tr}~g^{\mu\nu}F^{\alpha\beta}F_{\alpha\beta}.
\label{em1}
\ee
We note here that
the matrix model in the flat direction of long distance regime
could be obtained by dimensional reduction of
$D=10$ $\mbox{U(1)}^N$ super Yang-Mills theory to no space 
direction.
After dimensional reduction of $T^{\mu\nu}$ (\ref{em1})
associated with the broken $\mbox{U(1)}^N$ gauge symmetry,
all the components of $T^{\mu\nu}$
are reduced to the following nontrivial 
components;
$T^{00}$ and $T^{ij}$ alone.
The letters ``$i,~j$'' and ``$0$'' label 
the nine space directions and the time direction, respectively. 
Furthermore,
only one non-trivial conservation equation of $T^{\mu\nu}$
can survive;
\be
\partial_{0}~T^{00}=0,
\ee
where $T^{00}\sim 
\displaystyle{\sum_{i=1}^{9}}\mbox{Tr}~(\dot{A}^i)^2$.
We then find that
$\displaystyle{\sum_{i=1}^{9}}\mbox{Tr}~(\dot{A}^i)^2$ must
be a constant number.
This amounts to the dimensionally reduced $T^{\mu}_{\enskip\mu}$.
Here we must recall that the trace $T^{\mu}_{\enskip\mu}$ can be 
a non-zero value in contrust to the case of four dimensions.
Thus, the ninth component $\mbox{Tr}~(\dot{A}^9)^2$
can be obtained in terms of eight other components,
so that only the transverse eight components are left
as independent components;
\be
\mbox{Tr}~(\dot{A}^9)^2=~\mbox{const.}-\displaystyle{ 
 \sum_{i=1}^{8} }(\dot{A}^i)^2.
\ee
In the matrix model, $A^i$ is regarded as the position
coordinate $X^i$ of $N$ bounded D0-particles.
Thus, it is then possible to say that the bosonic matrix coordinate
$X^i$ has actually eight components rather than nine.

Moreover, we suppose that the action of the matrix model
describing a system of
low energy $N$ D0-particles
in large distance regime at weak string coupling
could be preserved
further in the large $R_{11}$ limit,
we can see that the action (\ref{matrixaction}) actually
corresponds to the matrix model in the flat direction of
the large distance regime at strong string coupling.
Henceforce, we can also identify the coupling $\lambda$ 
(\ref{lambda})
with the D0-particle mass $1/l_{s}g_{s}$;
\be
\frac{p^+ R_{11}}{l_{s}g_{s}}~\sim~\frac{1}{l_{s}g_{s}},
\ee
and we have the relation with the string tension
$T_{s}$
on the assumption that
the wrapped supermembrane after taking the limit 
$R_{11}\rightarrow \infty$
must be identified with the type IIA string
as in case of \cite{duff};
$T_{s}=2\pi R_{11}T_{SM}$ (\ref{m-s}).
We therefore obtain the following relation;
\be
p^+~\sim~\frac{1}{l_{s}g_{s}}.
\ee
Thus, the light-cone momentum $p^+$ becomes zero
in the large $R_{11}$ limit.
Therefore, the momentum of the only 8 space directions can be
thought to have non-zero value.
The theory turns out to be free to set
the 9-11 flip in the space directions in this limit as expected.

\section{\rm Discussions}
In this paper, it is shown that
the matrix model in the flat direction of large distance regime
can be directly connected with
the 11$D$ theory of 256 degrees of freedom
in the transverse light-cone frame
by taking the limit $R_{11}\rightarrow \infty$.
First, let us consider the meaning of this large $R_{11}$ limit
in the matrix model.
By taking account of the relations
$R_{11}\sim g^{2/3}_{s}l_p\sim l_{s}^{-2}l_p^3$, 
it seems that the large $R_{11}$ limit leads us to both
the strong string coupling and short string scale region.
The constant $l_p$ denotes the
11 dimensional Planck length.
This means that the string sector is restricted to
some local regions in the space-time.
We need not take account of any effects of the strings
in this limit.
Thus, it is possible to say that
the string sector substantially decouple from
the parton sector.
The situation seems to be the same one as in
the flat direction condition of the large distance regime. 
Therefore, the fact shows that the large $R_{11}$ limit yields
the flat direction in the matrix model.
On the other hand,
we have shown that the center of mass of the type IIA string
obtained from the 11$D$ supermembrane
in the large $R_{11}$ limit has the same matrix description
as the BFSS model.
Taking these matters into consideration,
we can see that
the two matrix descriptions are consistent
with each other under
the limit $R_{11}\rightarrow \infty$.

We would next like to comment on
the large $N$ limit with the large $R_{11}$ limit.
In this limit $R_{11}\rightarrow \infty$,
the energy of the whole system of the matrix model
has infinite energy.
In order to obtain the finite energy of the whole system
at strong string coupling,
it is necessary to take the limit
$N\rightarrow \infty$,
so that the energy of each state vanishes like $1/N$ \cite{bfss}.
Taking this large $N$ limit is equivalent to
treating the whole Fourier modes.
Thus, the energy of the matrix description derived
from 11$D$ supermembrane in the previous section
naturally remain finite in spite of the strong string coupling.

Finally, let us make a remark about the 11th direction.
This direction is a crucial one
to bring the longitudinal momentum $p^{11}$ 
to the 10$D$ theory as a nonperturbative R-R charge.
In the scheme of the so-called matrix string theory \cite{dvv-bs},
the role of the 11th direction
is thought to be different from other direction which leads us
to get a type IIA string, e.g., $X^{9}$.
We can see that in this paper,
the different roles of these directions belong to only one
of all directions in the strong string coupling,
i.e., the 11th direction $X^{11}$.
In the weak coupling region,
the situation is different from the case
in the strong string coupling ?
Our inference is that the situation have no change.
But we need to have an appropriate interpretation
about the KK nonzero-modes which are very massive
in turn, while the zero-mode corresponds to
a single IIA string \cite{duff}.
It seems undeniable that the KK nonzero-modes
correspond to the matrix model in the small $R_{11}$ limit.
In fact, we can see that
the correspondence with the zero-mode
is not mentioned in the matrix theory.\\

\vspace{0.6cm}

\noindent
{\it Acknowledgment}:
The author would like to thank Professor 
S.Saito for helpful discussions and careful reading
of the manuscript.


\end{document}